\documentclass[preprint,showpacs, superscriptaddress, preprintnumbers,amsmath,amssymb]{revtex4}

\usepackage{booktabs}
\usepackage{color,graphicx}
\usepackage{xcolor}
\usepackage{amsmath}
\usepackage{amssymb}
\usepackage{dcolumn}
\usepackage{bm}                  
\usepackage{soul}
\usepackage{enumerate}
\usepackage{tabularx}
\usepackage{multirow}
\usepackage{threeparttable}

\begin{document}

\title{
	The origin of the spin-orbit and pseudospin-orbit splittings' evolution in neutron drop}

\author{Yinghui Ge}
\affiliation{Department of Physics, School of Science, Tianjin University, Tianjin, 300354, China}

\author{Ying Zhang}
\email{yzhangjcnp@tju.edu.cn}
\affiliation{Department of Physics, School of Science, Tianjin University, Tianjin, 300354, China}
\affiliation{Strangeness Nuclear Physics Laboratory, RIKEN Nishina Center, Wako, 351-0198, Japan}

\author{Jinniu Hu}
\affiliation{School of Physics, Nankai University, Tianjin 300071, China}
\affiliation{Strangeness Nuclear Physics Laboratory, RIKEN Nishina Center, Wako, 351-0198, Japan}

\begin{abstract}
	We found the same staggering pattern in the evolution of the spin-orbit (SO) and pseudospin-orbit (PSO) splittings in neutron drops obtained by the energy-density functional (EDF) theory using SAMi-T and SLy5 parameterizations, with and without the tensor terms respectively.  This staggering evolution is 
    similar to the results recently obtained by the relativistic Brueckner-Hartree-Fock (RBHF) theory with Bonn A potential, which was claimed to be a tensor effect.  In this work, we present an 
    intuitive but essential explanation on the origin of this staggering evolution in the Skyrme EDF theory.  We found that for both the SO and PSO partners, the energy splittings mainly originate from the SO potential.   The staggering evolution of both the SO and PSO splittings is mainly due to the drastic change of density while neutrons filling different single-particle orbits.
    The key to determine the specific staggering pattern is the strength of the neutron density gradient term and the SO density term in the SO potential, not necessarily the tensor force.
    

\end{abstract}

\date{\today}

\pacs{
	21.10.Dr,  
	21.10.Pc,  
	21.60.Jz.  
}

\maketitle


A neutron drop is a system composed only of neutrons trapped by artificial external fields~\cite{Pudliner1996PRL}.  This simple system can be easily calculated by both the {\it ab initio} approach and the energy-density functional (EDF) theory. Therefore, the {\it ab initio} solutions for a neutron drop can be used as pseudo-data to calibrate and improve the effective Hamiltonians and density functionals in nuclear physics~\cite{Gandolfi2011PRL, Bogner2011PRC}.  Furthermore, the neutron drop can be also used to simulate the neutron-rich nuclei where the effect of protons could be replaced by an external field for simplicity~\cite{Chang2004NPA, Pieper2005NPA}.  

So far, the {\it ab initio} approaches used to investigate neutron drops mainly include the quantum Monte Carlo method~\cite{Pudliner1996PRL, Gandolfi2011PRL,Chang2004NPA,Pieper2005NPA, Smerzi1997PRC,Pieper2001PRC, Maris2013PRC,Carlson2015RMP,Pederiva2004NPA,Tews2016PRC,Zhao2016PRC}, the no-core shell model~\cite{Potter2014PLB}, and  the relativistic Brueckner--Hartree--Fock theory (RBHF)~\cite{Shen2018PLB,Shen2018PRC} with different realistic nucleon-nucleon potentials.  
These {\it ab initio} calculations with only two-nucleon (2N) potentials obtained similar total energies of neutron drops as a benchmark test.  However, there are still large uncertainties in the results calculated with different three-nucleon (3N) potentials for the neutron drops $N\gtrsim20$~\cite{ Maris2013PRC, Potter2014PLB}.  Compared with the {\it ab initio} solutions, the EDF theory with traditional Skyrme functionals~\cite{Pudliner1996PRL, Gandolfi2011PRL, Pederiva2004NPA, Bonnard2018PRC} usually provides smaller neutron drop total energies, but larger central densities and spin-orbit (SO) splittings. Based on this, some adjusted or new functionals have been worked out to achieve better agreements with the {\it ab initio} results~\cite{Pudliner1996PRL, Gandolfi2011PRL,Bonnard2018PRC,Kortelainen2014PRC}. 
Recently, the neutron drops have also been investigated by the covariant EDF theory with 
the latest and the adjusted relativistic effective interactions~\cite{Zhao2016PRC, Shen2018PLB, Shen2018PRC, Wang2019PRC}. 

In particular, Refs.~\cite{Shen2018PLB, Shen2018PRC} found a specific pattern in the evolution of SO and pseudospin-orbit (PSO) splittings in neutron drops.  By comparing with the {\it ab initio} results given by the RBHF theory and the results given by the covariant EDF theory with and without tensor contributions, they claimed that this specific pattern was due to the tensor effects.  Based on this, they put forward a new Skyrme functional SAMi-T with tensor terms to reproduce the neutron drop results given by the RBHF theory~\cite{Shen2019PRC}.  
Moreover, it was found in Ref.~\cite{Wang2019PRC} that to reproduce the RBHF results in the covariant EDF theory, the optimized tensor-force strength should vary with the strength of the external fields of neutron drops. 
Indeed, the tensor force plays a very sophisticated role on the evolution of the shell structure~\cite{Otsuka2005PRL}, which has been discussed extensively in the study of spin and pseudospin symmetry in ordinary nuclei (see Ref.~\cite{Liang2015PR} and the references therein).  However, it is well known that the effects of tensor force should be small due to the absence of the triplet $S$ channel in the pure neutron matter~\cite{Hu2010PLB}.  Recently, we found a similar evolution pattern of SO and PSO splittings in neutron drops calculated by the traditional Skyrme EDF theory without any tensor terms~\cite{Ge2020SC}.   
In this work, we will reveal the origin of this specific pattern of evolution. 


For simplicity and comparison, we suppose a spherical symmetry for the neutron drop, and neglect the pairing correlations among the neutrons as it was done in Refs.~\cite{Shen2018PLB, Shen2018PRC}. The wave function of a given single-particle state $i\equiv njl$ can be written as $\varphi_{i}(\mbox{\boldmath$r$}\sigma) = \frac{R_i(r)}{r}Y_{ljm}(\hat{r}\sigma)$,
where $R_i(r)$ is the radial wave function, and $Y_{ljm}(\hat{r}\sigma)$ is the spin spherical harmonics.
With the Skyrme functional, the radial Schr\"{o}dinger equation for the single-particle state $i$ with the energy $\varepsilon_{i}$ is 
\begin{eqnarray}
&&  -\frac{\hbar^2}{2m^*_q}\frac{d^2R_{i}}{dr^2} - \frac{d}{dr}\left( \frac{\hbar^2}{2m^*_q}\right) \frac{dR_{i}}{dr} + \left\{ U_q(r) + \frac{\hbar^2}{2m^*_q}\frac{l(l+1)}{r^2} 
+ \frac{1}{r}\frac{d}{dr}\left( \frac{\hbar^2}{2m^*_q}\right)  \right. \nonumber \\
&&  \left. +\left[ j(j+1)-l(l+1)-\frac{3}{4}\right] \frac{1}{r}W_q(r) \right\} R_{i}= \varepsilon_{i}R_{i}. \label{eq:Schrodinger}
\end{eqnarray}
The detailed expressions of the effective mass $m^*_q$ and the central Hartree-Fock potential $U_q(r)$ for $q=n,~p$ can be found in many literatures such as Ref.~\cite{Dobaczewski1984NPA}.  For a neutron drop, an external harmonic oscillator (HO) field $V_{\rm ext} = \frac{1}{2}m_n\omega^2 r^2$,
with $\frac{\hbar^2}{m_n}=41.44$~MeV fm$^2$ and $\hbar\omega=10$~MeV
is added in $U_n(r)$ to trap the neutrons as in Ref.~\cite{Shen2018PLB, Shen2018PRC}.   Besides, the center-of-mass correction is not included in the total energy, since the translation symmetry is lost due to the local external field.

Including the tensor terms of the Skyrme interaction, the Hamiltonian that will contribute to the SO potential can be written as
\begin{equation}
\mathcal{H}_{J}= 
\frac{1}{2}W_0 \left[J \rho'+
\sum_{q}cJ_q \rho'_q\right] 
+ \frac{1}{2} \alpha \sum_{q} J^2_q + \beta J_n J_p,
\end{equation}
where $\rho'~(\rho'_q)$ is the first-order derivative of the total (neutron/proton) density, and $J~(J_q)$ is the total (neutron/proton) SO density.  The coefficients ${\alpha=\alpha_C+\alpha_T}$ and ${\beta=\beta_C+\beta_T}$ are contributed from the central exchange term ${\alpha_C= \frac{1}{8}(t_1-t_2-t_1x_1-t_2x_2)}$, ${\beta_C = -\frac{1}{8}(t_1x_1+t_2x_2)}$, and the tensor term ${\alpha_T = \frac{5}{12}U}$, ${\beta_T = \frac{5}{24}(T+U)}$ respectively~\cite{Colo2007PLB}.  For a neutron drop, the obtained SO form-factor is 
\begin{equation}
W_n(r) 
= \frac{1}{2}W_0 (1+c)\rho'_n+  \alpha J_n. 
\label{eq:Wn}
\end{equation}  
In the following calculation, we choose two parameter sets: (1) SLy5 without tensor terms~\cite{Chabanat1998NPA}, where ${W_0=126.0}$~MeV~fm$^5$, $c=1$, and ${\alpha=\alpha_C=80.2}$~MeV~fm$^5$; (2) SAMi-T with tensor terms, which was newly fitted to reproduce the RBHF results for neutron drops~\cite{Shen2019PRC}, where ${W_0=130.0}$~MeV~fm$^5$, ${c=W'_0/W_0=0.78}$, ${\alpha_C=112.8}$~MeV~fm$^5$, $\alpha_T=-39.8$~MeV~fm$^5$, and thus ${\alpha=73.0}$~MeV~fm$^5$. 
One should notice that, the values of $W_0$ and $\alpha$ in SLy5 and SAMi-T are actually very close.  

\begin{figure}[h]
	\centering
	\includegraphics[width=0.5\textwidth]{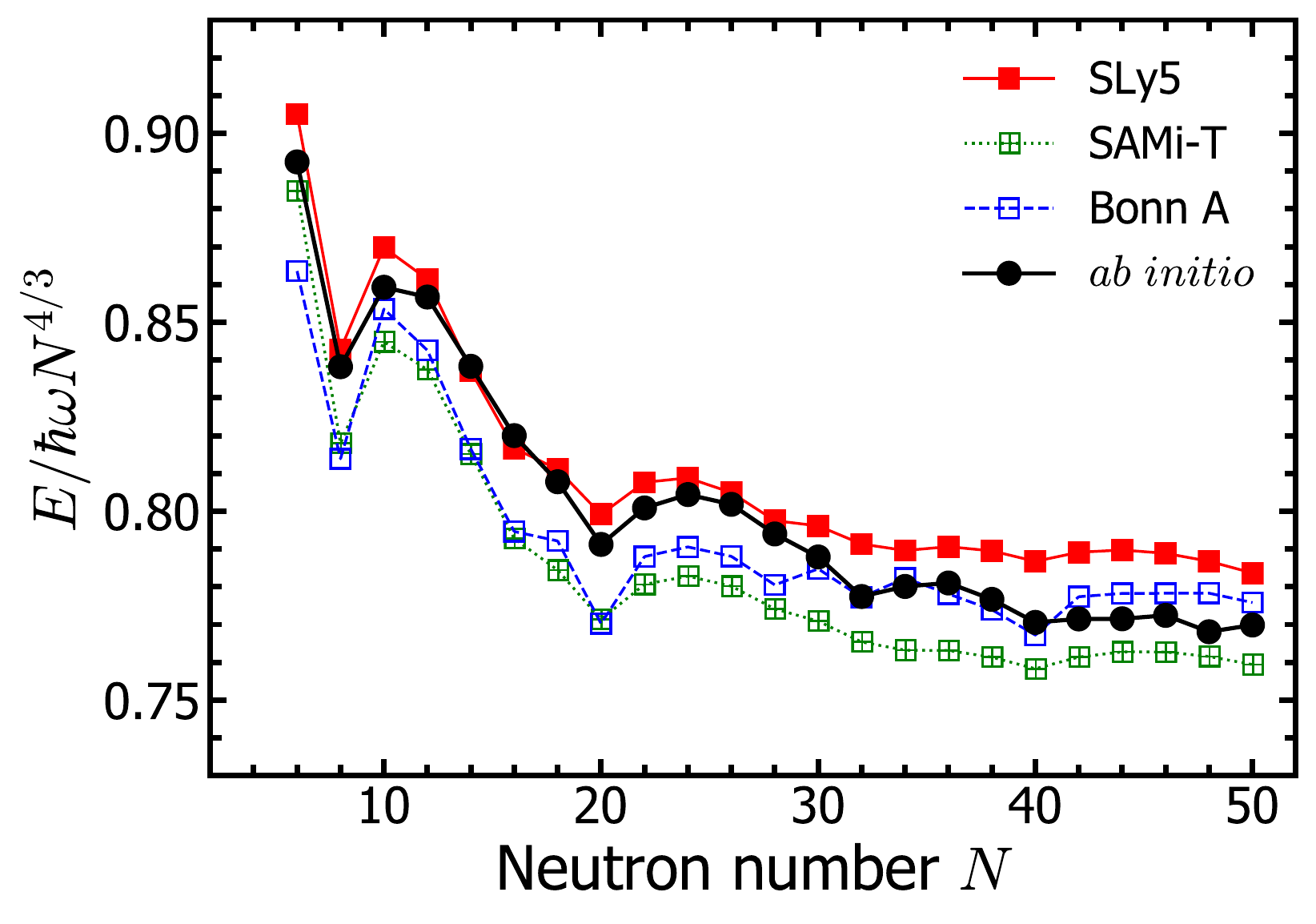}\\
	\caption{Total energies of neutron drops $N=6-50$ scaled by $\hbar\omega N^{4/3}$, calculated by EDF theory with SLy5, SAMi-T and RBHF theory with Bonn A~\cite{Shen2018PLB}.  The average of several other {\it ab initio} results are taken from Ref.~\cite{Bonnard2018PRC}.}\label{fig:SLy5-BonnA-HF-E}
\end{figure}

First, we present the total energies $E$ of neutron drops $N=6-50$ scaled by  the factor $\hbar\omega N^{4/3}$~\cite{Gandolfi2011PRL} calculated by the EDF theory with SLy5 and SAMi-T in Fig.~\ref{fig:SLy5-BonnA-HF-E}.   As a comparison, we also plot the results obtained by the RBHF theory calculated with Bonn A potential~\cite{Shen2018PLB,Shen2018PRC}, and the average result of several {\it ab initio} calculations~\cite{Bonnard2018PRC} including the quantum Monte Carlo method, the shell model, and the coupled-cluster calculations using different 2N interactions. The SLy5 results are rather close to the average {\it ab initio} results in neutron drops $N<30$, even better than the Bonn A results.  
The SAMi-T results are close to the Bonn A results only in $N<28$.  Respectively, SLy5 and SAMi-T provide the largest and smallest total energies for neutron drops $N>20$.

\begin{figure}[h]
	\centering
	\includegraphics[width=0.5\textwidth]{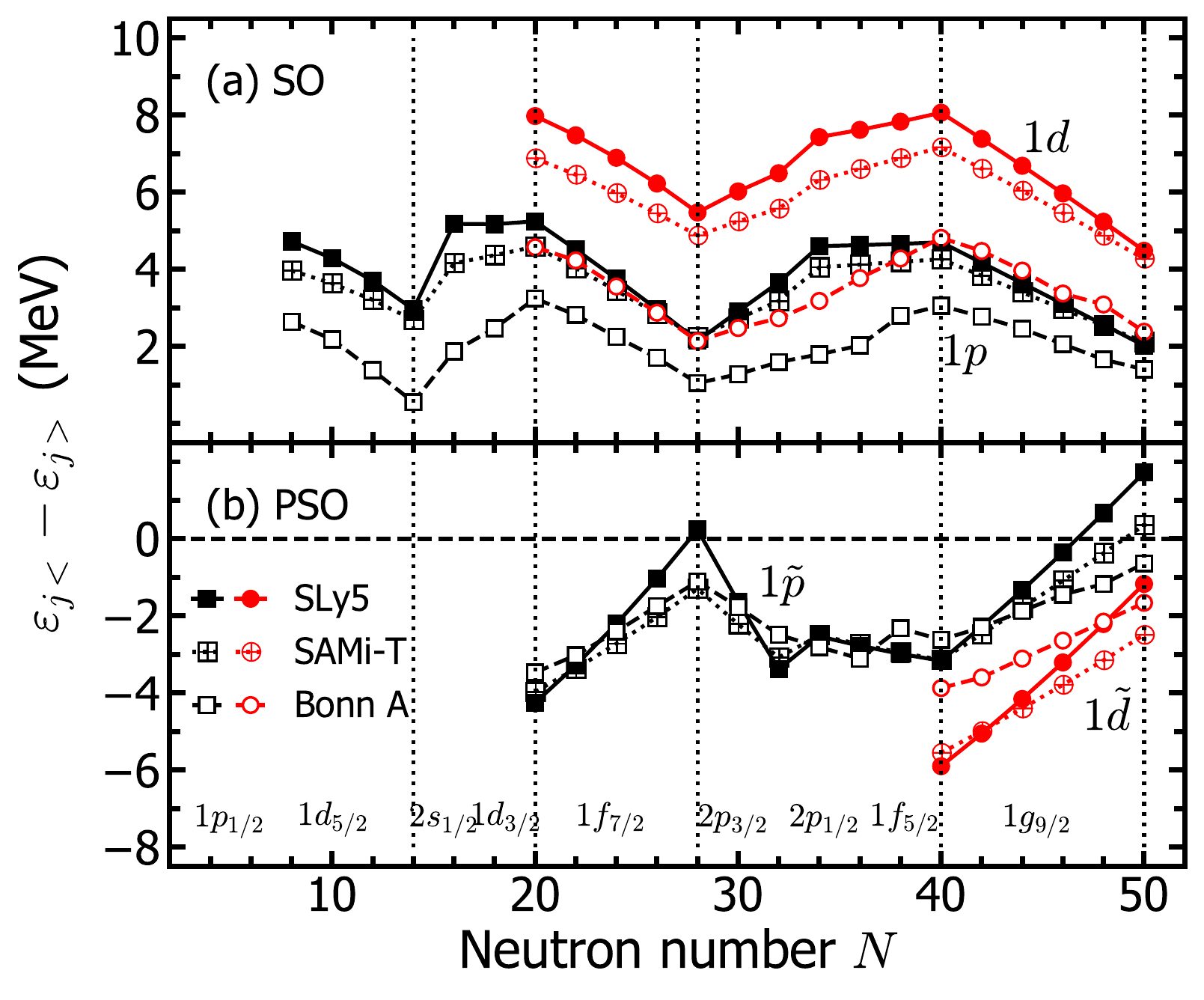}\\
	\caption{Spin-orbit (SO) splittings of $1p, 1d$ partners (a) and pseudospin-orbit (PSO) splittings of $1\tilde{p}, 1\tilde{d}$ partners (b) in neutron drops $N=6-50$ obtained by EDF theory with SLy5, SAMi-T and RBHF theory Bonn A~\cite{Shen2018PRC}. The labels ‘$1p_{1/2}, 1d_{5/2},$ etc.’ at the bottom denote the single-particle states occupied in order.  }\label{fig:SOS-PSOS}
\end{figure}


Next, we plot the evolution of the energy splittings ${\varepsilon_{j<}-\varepsilon_{j>}}$ between the SO partners, such as ${1p\equiv(1p_{1/2}, 1p_{3/2})}$, ${1d\equiv(1d_{3/2}, 1d_{5/2})}$, obtained by SLy5, SAMi-T and Bonn A potentials~\cite{Shen2018PLB, Shen2018PRC} as a function of the neutron number in Fig.~\ref{fig:SOS-PSOS} (a).  
The SLy5 without tensor and SAMi-T with tensor give quite similar SO splittings to each other, but obviously larger than the  Bonn A results.  However,  all the three results show a similar staggering pattern in the evolution as the neutron number increases.  Namely, the SO splittings reach the local minimum at $N=14, 28$, and the local maximum at $N=20, 40$ for both $1p$ and $1d$ partners. 
Meanwhile, the PSO splittings ${\varepsilon_{j<}-\varepsilon_{j>}}$ between the two partners, such as ${1\tilde{p}\equiv (2s_{1/2}, 1d_{3/2})}$ and ${1\tilde{d}\equiv (2p_{3/2}, 1f_{5/2})}$ in Fig.~\ref{fig:SOS-PSOS} (b) show an `opposite' staggering pattern to that of the SO splittings.  The negative value of the PSO splitting means that the level $\varepsilon_{j<}$ is lower than its partner $\varepsilon_{j>}$ in some neutron drops.  Therefore, the staggering pattern of the PSO splitting's evolution is actually similar to that of the SO splitting.  Such evolution  was explained by the tensor effect in Ref.~\cite{Shen2018PRC}.   Based on this, SAMi-T with tensor terms was fitted~\cite{Shen2019PRC}. However, in SLy5 there is no tensor term. 

In order to understand the origin of this specific staggering evolution found in SLy5, we separate the single-particle energy for a given state $i$ obtained by the Schr\"{o}dinger equation (\ref{eq:Schrodinger}) into three parts, 
$\varepsilon_i=\varepsilon_{1i} + \varepsilon_{2i}+\varepsilon_{3i}$, where
\begin{subequations}
	\begin{eqnarray}
	\varepsilon_{1i} &=&  \int^\infty_0 \left[ - \frac{\hbar^2}{2m^*_n}\frac{d^2R_{i}(r)}{dr^2} - \frac{d}{dr} \left( \frac{\hbar^2}{2m^*_n}\right) \frac{dR_{i}(r)}{dr} \right] R_{i} dr,  \\
	\varepsilon_{2i} &=&  \int^\infty_0  \left[ U_n(r) + \frac{\hbar^2}{2m^*_n}\frac{l(l+1)}{r^2} 
	+ \frac{1}{r}\frac{d}{dr}\left( \frac{\hbar^2}{2m^*_n}\right)
	\right]  R^2_{i}dr, \\
	\varepsilon_{3i} &=&  \int^\infty_0 \left[ j(j+1)-l(l+1)-\frac{3}{4}\right]  \frac{1}{r}W_n(r) R^2_{i} dr, 
	\end{eqnarray} 	
\end{subequations}
which are contributed from the derivatives of the single-particle wave functions, the HF and the centrifugal potentials, and the SO potential respectively.  
Therefore, the energy splitting between two single-particle states $i$ and $j$ can be also separated into three parts, 
${\Delta\varepsilon= \varepsilon_i-\varepsilon_j= \Delta\varepsilon_{1} + \Delta\varepsilon_{2}+\Delta\varepsilon_{3}}$.  

For a pair of SO partners, such as ${1p=(1p_{1/2}, 1p_{3/2})}$, their wave functions $R_i$ are quite similar to each other.  So it is easy to imagine that, in a given nucleus all the three parts ${\Delta\varepsilon_{1}, \Delta\varepsilon_{2}, \Delta\varepsilon_{3}}$ of the energy splitting between the SO partners should be very small.  That's why one could see the approximate spin symmetry among the single-particle levels in the ordinary nucleus.  
While for a pair of PSO partners, such as ${1\tilde{p}=(2s_{1/2}, 1d_{3/2})}$, their  wave functions $R_i$ are quite different, since their numbers of radial nodes and the angular momentum are totally different. Therefore, their energy differences ${\Delta\varepsilon_{1}, \Delta\varepsilon_{2}, \Delta\varepsilon_{3}}$ could be very large.
In spite of this, their total energy splitting $\Delta \varepsilon$ turns out to be rather small, since one could also see the approximate pseudospin symmetry among the single-particle levels in the ordinary nucleus, sometimes even better than the spin symmetry.  This indicates that there must be some cancellation among the three parts ${\Delta\varepsilon_{1}, \Delta\varepsilon_{2}, \Delta\varepsilon_{3}}$.

\begin{table}[htbp]
	\caption{The total energy splitting $\Delta\varepsilon$ between the SO partners $1p, 1d$,  the PSO partners $1\tilde{p}, 1\tilde{d}$, and its three parts of contribution $\Delta\varepsilon_{1}, \Delta\varepsilon_{2}, \Delta\varepsilon_{3}$ in neutron drops $N=20, 28, 40, 50$, calculated by the EDF theory with SLy5.  The units are MeV.  }
	\begin{tabular}{cc|rrr|r}
		\hline 
		&      &  {\centering$\Delta \varepsilon_1$}     & {\centering$\Delta \varepsilon_2$}     &  {\centering$\Delta \varepsilon_3$}    &   {\centering$\Delta \varepsilon$ }  \\ \hline
		\multirow{3}{*}{\centering $N=20$}	
		&  $1p$              &  $-0.203 $     &  $0.211$    &   $5.233 $   &  $5.241 $    \\
		&  $1d$              &   $-1.251$     &  $1.313 $    &  $7.912 $    &   $7.974 $   \\
		&  $1\tilde{p}$   &   $20.976 $   &  $-20.590 $    &  $-4.639 $    &   $-4.253 $   \\ \hline
		\multirow{3}{*}{\centering $N=28$}	
		&  $1p$              &  $0.750 $    &  $-0.712 $    &  $2.104 $    &   $2.142 $   \\
		&  $1d$              &  $-0.189 $    &  $0.238 $    &  $5.422 $    &   $5.471 $   \\
		&  $1\tilde{p}$   &  $22.782 $    &  $-19.394 $    &  $-3.148 $    &   $0.240 $   \\ \hline
		\multirow{4}{*}{\centering $N=40$}	
		&  $1p$              &  $-0.058 $    &  $0.062 $    &  $4.694 $    &   $4.698 $   \\
		&  $1d$              &  $-0.344 $    &  $0.351 $    &  $8.054 $    &   $8.061 $   \\
		&  $1\tilde{p}$   &  $23.379 $    &  $-21.725 $    &  $-4.818 $    &   $-3.164 $   \\
		&  $1\tilde{d}$   &  $19.076 $    &  $-17.636 $    &  $-7.340 $    &   $-5.900 $   \\ \hline
		\multirow{4}{*}{\centering $N=50$}	
		&  $1p$              &  $0.436 $    &  $-0.423 $    &  $2.003 $    &   $2.016 $   \\
		&  $1d$              &  $0.618 $    &  $-0.590 $    &  $4.450 $    &   $4.478 $   \\
		&  $1\tilde{p}$   &  $24.696 $    &  $-20.375 $    &  $-2.598 $    &   $1.723 $   \\
		&  $1\tilde{d}$   &  $19.844 $    &  $-15.622 $    &  $-5.392 $    &   $-1.170 $  \\
		\hline
	\end{tabular}\label{tab:esp123}
\end{table}

In Table~I, we list the three parts of contribution $\Delta\varepsilon_{1}, \Delta\varepsilon_{2}, \Delta\varepsilon_{3}$ and the total energy splitting $\Delta \varepsilon$ calculated by SLy5 for the SO partners $1p$, $1d$, and the PSO partners $1\tilde{p}$, $1\tilde{d}$ in neutron drops $N=20, 28, 40, 50$.  As we expected, for the SO splittings, the values of all the three parts are small, especially the first two parts $\Delta\varepsilon_{1}$ and $\Delta\varepsilon_{2}$.  Moreover, the first two parts have opposite signs and nearly cancel with each other.  As a result,  the total SO splitting $\Delta \varepsilon$ is dominated by the third part $\Delta\varepsilon_{3}$.  Indeed, this part is calculated by the SO potential which was originally introduced to reproduce the nuclear magic numbers by a large SO splitting for the high angular momentum states~\cite{Haxel1949PR, Mayer1949PR}.  
More interestingly, for the PSO splittings, although the contributions from $\Delta\varepsilon_{1}$ and $\Delta\varepsilon_{2}$ are quite large as we expected, they also have opposite signs and tend to cancel with each other.  Especially in neutron drop $N=20$, this cancellation is very obvious, and thus the total PSO splitting $\Delta \varepsilon$ is also dominated by the third part $\Delta\varepsilon_{3}$.  Although in the heavier neutron drops, the cancellation between $\Delta\varepsilon_{1}$ and $\Delta\varepsilon_{2}$ is not that obvious, the third part $\Delta\varepsilon_{3}$ is still important to determine the total energy splitting.   

Actually, it is easy to understand the approximate cancellation between the first two parts $\Delta\varepsilon_{1}$ and $\Delta\varepsilon_{2}$ in both the SO and PSO splittings.  One could recall the single-particle levels calculated only with the HO potential, which are degenerate within the major shell $N_d=2n_r+\ell$.  Namely, the $1p$ states have the same energy, $2s$ and $1d$ also have the same energy.  In this case, there is no $\Delta\varepsilon_{3}$ contribution to the total energy splitting at all.  Therefore, $\Delta\varepsilon_{1}$ and $\Delta\varepsilon_{2}$ have to cancel with each other exactly between any SO or PSO partners.  Now with the effective Skyrme interactions in addition, this cancellation is not exact, but still leads to the dominance of $\Delta\varepsilon_{3}$ in the total energy splitting.  

\begin{figure}[htbp]
	\centering
	\includegraphics[width=0.5\textwidth]{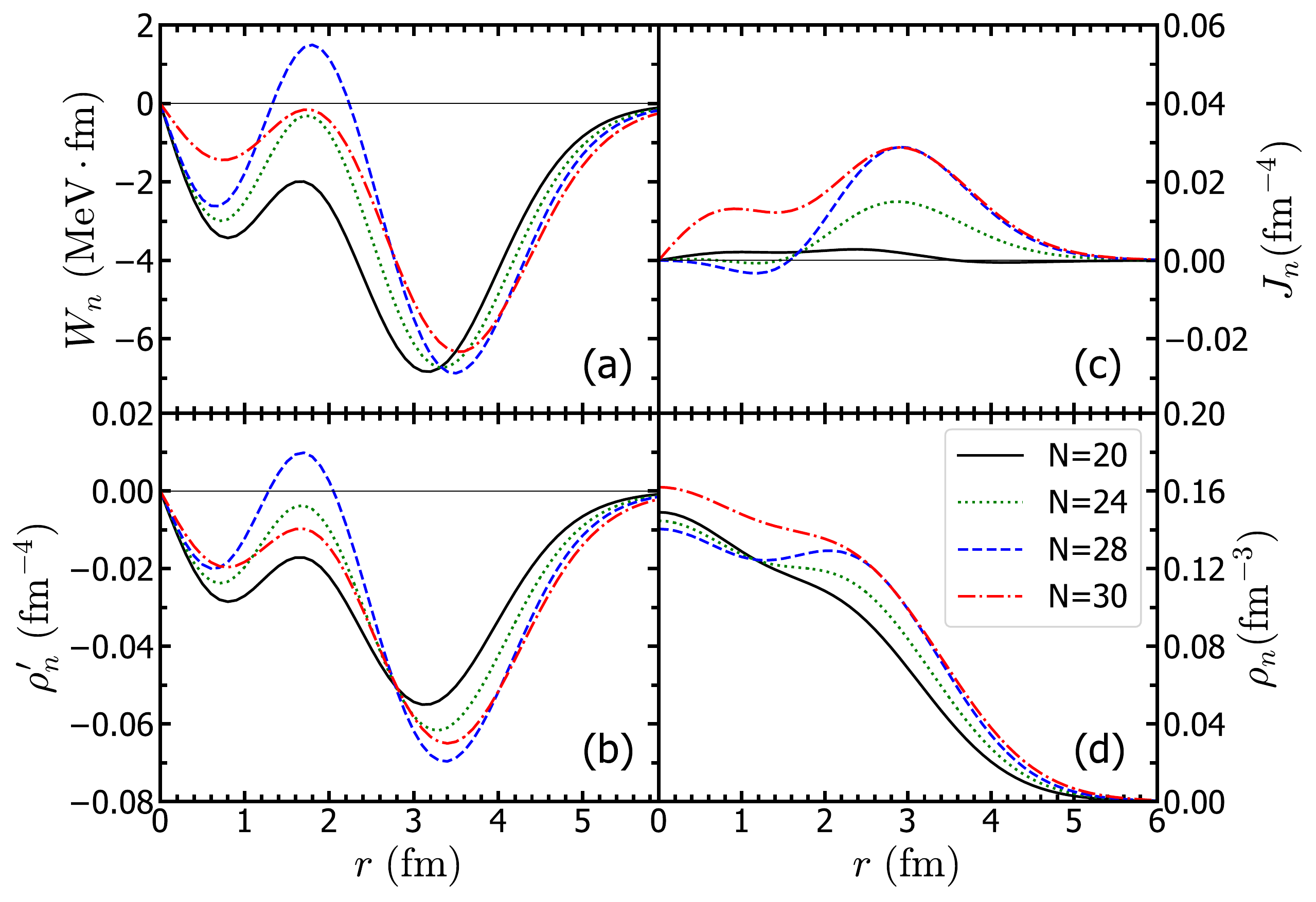}\\
	\caption{SO form-factor $W_n$ (a), first-order derivative of neutron density $\rho'_n$(b), SO density $J_n$ (c), and neutron density $\rho_n$ (d) for neutron drops $N=20, 24, 28, 30$ calculated by SLy5.  }\label{fig:W-rho-J}
\end{figure}

Then, we focus on the evolution of SO form-factor $W_n$, which is the key of the integral to calculate $\Delta\varepsilon_{3}$.   Figure~\ref{fig:W-rho-J}(a) shows the distribution of $W_n$ in neutron drops $N=20, 24, 28, 30$ calculated with SLy5.  
In these neutron drops, all the values of $W_n$ are negative except $N=28$ at $r\approx 2$~fm. 
More explicitly, from $N=20$ to $N=28$, the local maximum of $W_n$ at $r\approx 2$~fm is raised up obviously from the negative, up to be positive at $N=28$.  Then from $N=28$ to $N=30$, this local maximum drops back to be negative.  
Meanwhile the value of local minimum at $r\approx 3$~fm almost remains the same, but shifts further outside from $N=20$ to $N=30$.  Especially at $N=28$, the positive values of $W_n$ at $r\approx 2$~fm will cancel with parts of the negative ones in the integral to calculate $\Delta\varepsilon_{3}$, and these negative values are similar among different neutron drops.  As a result, $\Delta\varepsilon_{3}$ 
has the minimum absolute value at $N=28$.
This could explain the local minimum of the total SO splitting, and the local maximum of the total PSO splitting at $N=28$ found in Fig.~\ref{fig:SOS-PSOS}.  

As we wrote in Eq.~(\ref{eq:Wn}), the SO form-factor $W_n$ is contributed by two parts, i.e., the first-order derivative of the density $\rho'_n$ and the SO density $J_n$.  They are plot in Fig.~\ref{fig:W-rho-J} (b) and (c) respectively.  It is obvious to see that, the distribution of $W_n$ is generally determined by $\rho'_n$.  With a close look, we find that the local maximum and minimum of $\rho'_n$ are raised up and pushed down visibly at $r\approx 2$~fm and $r\approx 3$~fm respectively as neutron number increases.  At the same time, the positive SO density $J_n$ mainly contributes in the outer region at $r\approx 3$~fm, where the spin could not be saturated.   As a result,  $J_n$ helps to push the negative local minimum of $W_n$ at $r\approx 3$~fm back to the similar values for different neutron drops, but does not affect the local maximum at $r\approx 2$~fm so much.  Therefore, the shift up and down of the local maximum of $W_n$ at $r\approx 2$~fm is mainly determined by the change of $\rho'_n$.

To further understand the change of $\rho'_n$ at $r\approx 2$~fm, we plot the neutron density $\rho_n$ in Fig.~\ref{fig:W-rho-J}(d).   From $N=20$ to $N=28$, the neutron density $\rho_n$ grows intensively at $r\approx 2.5$~fm, which leads to an increase of $\rho'_n$ nearby.   Obviously, this is due to the successive filling of the $1f_{7/2}$ orbit as shown in Fig.~\ref{fig:SOS-PSOS}.  Especially at $N=28$, when the orbit $1f_{7/2}$ is fully occupied, the density at $r\approx 2.5$~fm increases to the maximum as a bump, and thus its first-order derivative $\rho'_n$ at  $r\approx 2$~fm turns out to be positive.  
At $N=30$, the next two neutrons come to the $2p_{3/2}$ orbit as shown in Fig.~\ref{fig:SOS-PSOS}.  Then the central neutron density grows obviously at $r\lesssim 2$~fm due to the contribution from the single-particle wave function of $2p_{3/2}$.  As a result, $\rho'_n$ at $r\approx 2$~fm becomes negative again. 

From the above analysis, we can understand that the staggering evolution of the SO splitting for $1p, 1d$ partners and the PSO splitting for $1\tilde{p}, 1\tilde{d}$ partners in neutron drops around $N=28$, is mainly due to the drastic change of density while neutron filling the $1f_{7/2}$ orbit first and then the $2p_{3/2}$ orbit.  This could also help us to understand the staggering evolution in other neutron drops.  For example, the decrease of the SO splitting for $1p$ partners from $N=8-14$ is due to the filling of the $1d_{5/2}$ orbit, then the increase from $N=16$ is due to the filling of the $2s_{1/2}$ orbit.  Similarly, the decrease of the SO splitting and the increase of PSO splitting from $N=40$ are due to the filling of the $1g_{9/2}$ orbit.  
This is because
the larger angular momentum state contributes to the total density most in the outer region, while the lower angular momentum state most in the inner region.   
The drastic change of density due to the step-filling of neutrons in different orbits leads to the shift up and down of the SO potential near the neutron drop surface, and thus gives rise to the staggering evolution of the SO and PSO splittings.  

The above analysis and conclusions are also valid for the SAMi-T results. Now, we can understand why a rather similar SO and PSO splittings could be found in Fig.~\ref{fig:SOS-PSOS} between SLy5 and SAMi-T.  This is just because their coefficients $W_0$ before $\rho'_n$ and $\alpha$ before $J_n$ in the SO form-factor $W_n$ are quite similar to each other.


In summary, by using the Skyrme EDF theory with SLy5 and SAMi-T parameterizations, we obtained consistent results with several {\it ab initio} calculations for neutron drops $N=6-50$.  No matter the tensor terms are included or not, 
 we found the same staggering pattern in the evolution of the SO and PSO splittings in neutron drops obtained by SAMi-T and SLy5, which was similar to the results obtained by the RBHF theory with Bonn A potential.  We presented an explanation on the origin of this staggering evolution.  By separating the energy splittings into three parts, we found that the part calculated by the SO potential dominates the total energy splitting between both the SO and PSO partners.  Further investigation on the evolution of this SO potential and the density shows that, the staggering evolution of both the SO and PSO splittings is mainly due to the drastic change of density while neutron filling different single-particle orbits.  The key to determine the specific staggering pattern is the strength of the neutron density gradient term and the SO density term in the SO potential, not necessarily the tensor force.

Here, one may notice that the contribution from the SO density to the SO form-factor  plays a delicate role in the staggering evolution.  We have also tried SLy4 and SkM*~\cite{Ge2020SC}, which does not have this SO density contribution, but obtained a similar staggering evolution.  
Furthermore, the pairing correlation is neglected in the present investigation for the comparison with the RBHF results.  One could expect that the scattered occupation around the Fermi energy due to the pairing correlation will lead to a more gentle change of the density, and thus a smoother staggering evolution of the SO and PSO splittings~\cite{Ge2020SC}.  

The authors are grateful to H. Z. Liang, S. H. Shen and H. Sagawa for the valuable discussions during this work. 
This work was partly supported by the National Natural Science Foundation of China (NSFC) 
(Grants No. 11405116, No. 11775119),  and China Scholarship Council (Grant No. 201906255002 and No. 
201906205013).

\clearpage

\end{document}